# Quantum Teleportation of Single Qubit Mixed Information State with Werner-Like State as Resource

Hari Prakash[*,1,2] and Vikram Verma[**,1]

[1]Physics Department, University of Allahabad, Allahabad-211002, India.
[2]Indian Institute of Information Technology, Allahabad-211012, India.

E-mail: [*]prakash_hari123@rediffmail.com
[**]vikramverma18@gmail.com

**Abstract:** In this paper we extend our recent results for quantum teleportation of single qubit pure information state using non-maximally entangled pure state resource [*Quantum Inf Process*, **11** (2012) 1251] to the case of a general mixed information state and mixed Werner states resource. Fidelity is found dependent on information state in general. We minimize it over all possible information states and maximize it over possible unitary transformations which can be done by the receiver to obtain a value dependent on purity of information state and mixing parameter of Werner states. For pure information state, this value of fidelity is independent of information and for mixed information states it is less than unity even for perfect entangled resource. The latter result is in contrast to the case of pure states.

**Introduction:**

Quantum Entanglement is the main resource in Quantum Information Processing (QIP). Entangled system divided into two parts enables teleportation of quantum information of an unknown quantum state to a remote place while the original state is destroyed. In 1993, first time Bennett et al. [1] proposed a scheme for the quantum teleportation (QT) of an unknown quantum state of a single qubit from a sender (say, Alice) to a receiver (say, Bob) using an Einstein-Podolsky-Rosen (EPR) channel having perfectly entangled two qubit state and a classical channel. Two entangled qubits are shared between Alice and Bob. Alice performs Bell-State Measurement (BSM) on her two particles and communicates this BSM result to Bob through a classical channel. Bob performs a unitary transformation, dependent on result of BSM received from Alice and Bob's particles are converted to information state. This type of QT, known as standard quantum teleportation (SQT) is regarded as one of the most spectacular and interesting achievements in quantum theory due to its important applications in quantum computation and quantum communication. QT has been seen experimentally [2-4] by a number of authors. In all these experiments, however, the information is a single photon and standard bi-photonic entangled states (SBES) have been used. This however had a disadvantage in that complete BSM is not possible using only linear optics.

Another scheme using superposed coherent states (SCS) as information and entangled coherent states (ECS) as quantum resource was proposed by van Enk Hirota [5]. This had advantage of greater robustness against decoherence compared to the SBES. Van Enk and Hirota obtained fidelity dependent on information. The authors discussed average fidelity (average over information) and obtained success 1/2. Prakash et al. [6] modified this scheme for almost perfect success. These authors also used the concept of "Minimum Assured Fidelity (MASFI)" the minimum of fidelity against all possible information for information dependent fidelity.

SQT can be made more secure by involving other players in the game using entangled state of more than two qubits [7-13]. This has been known as controlled QT. SQT has also been generalized for teleportation of larger number of qubits [14-25].

Key to QT is the phenomenon of entanglement. If the quantum channel is maximally entangled, the quantum state is perfectly reproduced at Bob's location and the fidelity of the teleportation is unity. However, in the real world, the quantum channel lies in a noisy environment, which degrades entanglement of the channel. This is expected to reduce the fidelity.

The Bell states are maximally entangled states (MES). Hill and Wooters [26] defined a magic basis involving states differing from Bell states by a phase at the maximum. For an arbitrary non-maximally entangled state (NMES) as resource state, its concurrence [26-27] could be expressed in terms of sum of square of coefficients in its expansion in magic basis [26]. Use of magic basis and NMES for 3 qubits resource was studied by Prakash and Maurya [28]. The present authors [29] have shown that such a magic basis is not obtained for entangled states of 2N qubits with N>1.

Agrawal and Pati [30-31] studied probabilistic QT using NMES as resource. Prakash et.al.[32] used four partite NMES involving ECS for CQT of SCS information. The present authors [33] studied methodically use of pure NME states as resource in QT of pure single qubit information state and found that fidelity depends on information as well as the resource state in a complicated way. MASFI was seen to have a simple relationship, MASFI = 2C/(1+C), with concurrence C and clearly it increases monotonically with C. Behavior of minimum average fidelity (MAVFI) was however complicated. Prakash and Mishra [34] reported an interesting study of QT with NME coherent states and showed that substantialy increased MASFI and MAVFI could be obtained by reducing the concurrence.

Study of information and resource states as mixed state is more realistic as pure states get converted to mixed state by interaction with environment. In this paper we extend our recent results for pure information and resource states [33] to the case of mixed states. The theory of entanglement of mixed-state is more complicated and less well understood than that of pure state. We consider a Werner-Like mixed state as resource. Werner described a mixed state, called Werner state [35], which can be expressed as $\rho_W = \varepsilon |\psi^-\rangle\langle\psi^-| + \frac{1}{4}(1-\varepsilon)\mathrm{I}$, $0 \leq \varepsilon \leq 1$. It consists of a mixture of pure singlet state $|\psi^-\rangle = \frac{1}{\sqrt{2}}[|01\rangle - |10\rangle]$, a maximally entangled state, with probability ε ($0 \leq \varepsilon \leq 1$) and a fully mixed state with probability 1-ε. The concurrence of this Werner state is given by $C = \max\{0, (3\varepsilon - 1)/2\}$. The Werner state is pure only when $\varepsilon = 1$. Depending on the singlet weight ε, Werner states may be entangled ($\varepsilon > \frac{1}{3}$) or separable ($\varepsilon \leq \frac{1}{3}$). This definition can be generalized to include other states of two qubits [36-37] in place of the singlet state $|\psi^-\rangle$, giving $\rho_W = \varepsilon |M\rangle\langle M| + \frac{1}{4}(1-\varepsilon)\mathrm{I}$, $0 \leq \varepsilon \leq 1$, where $|M\rangle$ is any two-qubit maximally entangled states. These states are called Werner-like states and are important in quantum information theory.

**Quantum Teleportation of Mixed Single Qubit Information state using Werner State**

Consider most general single qubit information state given by

$$\rho_I = \begin{pmatrix} \rho_{00} & \rho_{01} \\ \rho_{10} & \rho_{11} \end{pmatrix}, \text{ with } \mathrm{Tr}(\rho_I) = \rho_{00} + \rho_{11} = 1, \ |\rho_{01}|^2 \leq \rho_{00}\rho_{11}. \tag{1}$$

It is convenient to define angles α and β and a parameter γ to write the elements of $\rho_I$ as

$$\rho_{00} = \cos^2\frac{\alpha}{2}, \ \rho_{11} = \sin^2\frac{\alpha}{2}, \ \rho_{01} = \gamma \sin\frac{\alpha}{2}\cos\frac{\alpha}{2}\exp(-i\beta), \ \rho_{01} = \gamma \sin\frac{\alpha}{2}\cos\frac{\alpha}{2}\exp(i\beta) \tag{2}$$

with $0 \leq \alpha \leq \pi$, $0 \leq \beta < 2\pi$ and $0 \leq \gamma \leq 1$. Parameter $\gamma$ defines the purity of the information state and the value $\gamma=1$ corresponds to pure information state. For arbitrary information state of a given purity one can vary α and β keeping γ constant.

Consider the Werner-like state [36-37] of two qubits is given by

$$\rho_W = \frac{1-\varepsilon}{4} I + \varepsilon |\phi^+\rangle\langle\phi^+|, \quad 0 \leq \varepsilon \leq 1 \tag{3}$$

where $|\phi^+\rangle = \frac{1}{\sqrt{2}}[|00\rangle + |11\rangle]$. The two qubits are found entangled for ε > 1/3 and the concurrence is given by $C = \max\{0, (3\varepsilon-1)/2\}$. Composite state containing qubit 1 of information state and qubits 2 & 3 in the Werner state can be described by the density operator

$$\rho_c^{(1,2,3)} = \rho_I^{(1)} \otimes \rho_W^{(2,3)}. \tag{4}$$

Let the first two qubits (1 & 2) go to the sender, Alice and the remaining qubit (3) to the receiver, Bob.

Later calculation of fidelity involving traces of products of density operators over states of given qubits become simple if one uses the operators

$$I_+ = \frac{1}{2}(I + \sigma_z) = |0\rangle\langle 0|, \quad I_- = \frac{1}{2}(I - \sigma_z) = |1\rangle\langle 1|,$$

$$R_+ = \frac{1}{2}(\sigma_x + i\sigma_y) = |0\rangle\langle 1|, \quad R_- = \frac{1}{2}(\sigma_x - i\sigma_y) = |1\rangle\langle 0|. \tag{5}$$

because the only nonzero traces of products of two such operators are

$$\text{Tr}[I_+ I_+] = \text{Tr}[I_- I_-] = \text{Tr}[R_+ R_-] = \text{Tr}[R_- R_+] = 1 \tag{6}$$

In terms of these operators, the information state of single qubit and resource Werner state of two qubits can be written as

$$\rho_I^{(1)} = \rho_{00} I_+ + \rho_{11} I_- + \rho_{01} R_+ + \rho_{10} R_- \tag{7}$$

$$\rho_W^{(2,3)} = \frac{1}{4}(1+\varepsilon)(I_+ I_+ + I_- I_-) + \frac{1}{4}(1-\varepsilon)(I_+ I_- + I_- I_+) + \frac{\varepsilon}{2}(R_+ R_+ + R_- R_-) \tag{8}$$

For Bell states, $|B_{0,1}\rangle = \frac{1}{\sqrt{2}}[|00\rangle \pm |11\rangle]$, $|B_{2,3}\rangle = \frac{1}{\sqrt{2}}[|00\rangle \pm |10\rangle]$ we can write similarly

$$\rho_{B_{0,1}} = |B_0\rangle\langle B_0| = \frac{1}{2}[I_+ I_+ + I_- I_- \pm R_+ R_+ \pm R_- R_-], \tag{9.a}$$

$$\rho_{B_2} = |B_2\rangle\langle B_2| = \frac{1}{2}[I_+ I_- + I_- I_+ \pm R_+ R_- \pm R_- R_+], \tag{9.b}$$

If Alice's BSM result is $|B_r\rangle$, then, the state that reaches to Bob will be

$$\rho_{(Bob)_r}^{(3)} = \mathrm{Tr}_{12}[\rho_I^{(1)} \rho_W^{(23)} \rho_{B_r}^{(12)}] \Big/ \mathrm{Tr}_{123}[\rho_I^{(1)} \rho_W^{(23)} \rho_{B_r}^{(12)}] \tag{10}$$

with probability $P_r = \mathrm{Tr}_{123}[\rho_I^{(1)} \rho_W^{(23)} \rho_{B_r}^{(12)}]$. Explicitly we obtain $P_r = 1/4$ and

$$\left.\begin{array}{l}
\rho_{(Bob)_0}^{(3)} = \frac{1}{2}[\{\rho_{00}(1+\varepsilon) + \rho_{11}(1-\varepsilon)\}I_+ + \{\rho_{00}(1-\varepsilon) + \rho_{11}(1+\varepsilon)\}I_- + 2\varepsilon\rho_{01} R_+ + 2\varepsilon\rho_{10} R_-] \\[4pt]
\rho_{(Bob)_1}^{(3)} = \frac{1}{2}[\{\rho_{00}(1+\varepsilon) + \rho_{11}(1-\varepsilon)\}I_+ + \{\rho_{00}(1-\varepsilon) + \rho_{11}(1+\varepsilon)\}I_- - 2\varepsilon\rho_{01} R_+ - 2\varepsilon\rho_{10} R_-], \\[4pt]
\rho_{(Bob)_2}^{(3)} = \frac{1}{2}[\{\rho_{00}(1-\varepsilon) + \rho_{11}(1+\varepsilon)\}I_+ + \{\rho_{00}(1+\varepsilon) + \rho_{11}(1-\varepsilon)\}I_- + 2\varepsilon\rho_{10} R_+ + 2\varepsilon\rho_{01} R_-] \\[4pt]
\rho_{(Bob)_3}^{(3)} = \frac{1}{2}[\{\rho_{00}(1-\varepsilon) + \rho_{11}(1+\varepsilon)\}I_+ + \{\rho_{00}(1+\varepsilon) + \rho_{11}(1-\varepsilon)\}I_- - 2\varepsilon\rho_{10} R_+ - 2\varepsilon\rho_{01} R_-].
\end{array}\right\} \quad 11)$$

It may be noted here that we can write $\rho_{(Bob)_r} = \sigma_r \rho_{(Bob)_0} \sigma_r^\dagger$ with $\sigma_r = I, \sigma_z, \sigma_x, i\sigma_y$ for r = 0, 1, 2, 3 respectively.

Dependent on result of Alice's BSM, $|B_r\rangle$, Bob performs a unitary transformation $U_r^{(3)}$ to replicate the information. The teleported state would be $\rho_T^{(3)} = U_r^{(3)} \rho_{(Bob)_r}^{(3)} U_r^{(3)\dagger}$. The fidelity for the case of BSM result r is then given by $F_r = Tr[\rho_T^{(3)} \rho_I^{(3)}]$. This fidelity is found dependent on the information state. Since there is no control over information we may consider the values of $F_r$ either (i) minimized or (ii) averaged over all possible information. The value of $F_r$ thus obtained should then be maximized over possible unitary transformation ($U_r$) which Bob can perform as he can choose it arbitrarily. It should be noted that this order of minimization and then maximization can be reversed only if the values of angles of $U_r$ at which the maximum occurs are found independent of angles α and β as there is no control over α and β. Since $\rho_{(Bob)_r} = \sigma_r \rho_{(Bob)_0} \sigma_r^\dagger$, it is clear that same teleported state $\rho_T^{(3)} = U_r^{(3)} \rho_{(Bob)_r}^{(3)} U_r^{(3)\dagger}$ is obtained if $U_r = U_0 \sigma_r$ is taken.

If we write

$$U_0 = e^{i\chi} \begin{pmatrix} \cos\frac{\theta}{2} e^{i\phi} & \sin\frac{\theta}{2} e^{i\psi} \\ -\sin\frac{\theta}{2} e^{-i\psi} & \cos\frac{\theta}{2} e^{-i\phi} \end{pmatrix} \quad (12)$$

with $0 \leq \theta \leq \pi$, $0 \leq \phi, \leq \pi, 0 \leq \psi \leq \pi$ and $0 \leq \chi < 2\pi$, the fidelity in each case of Alice's BSM results would be

$$F = \mathrm{Tr}[\rho_T \rho_I] = \frac{1}{2}[(1 + \varepsilon \cos\theta \cos^2\alpha) + \gamma\varepsilon \sin\theta \sin 2\alpha \sin\phi \sin(\beta + \psi)$$
$$+ \gamma^2 \varepsilon \cos^2\frac{\theta}{2} \cos 2\phi \sin^2\alpha - \gamma^2 \varepsilon \sin^2\frac{\theta}{2} \sin^2\alpha \cos 2(\beta + \psi)] \tag{13}$$

If we minimize F given in Eq.(15) against $\alpha$ & $\beta$ and maximize against $\theta, \phi, \psi$ we get

$$\mathrm{MASFI} = \frac{1}{2}[1 + \gamma^2 \varepsilon] \tag{14}$$

at $(\theta, \phi, \alpha) = (0, 0, \pi/2), (0, \pi, \pi/2)$. We note that the MASFI is not 1 even if $\varepsilon$=1 at which the Werner states are maximally entangled. This behavior is in contrast with the behavior of pure information states [33] which gives perfect QT if the quantum resource is maximally entangled. We also note that if $\theta = \phi = 0$ is taken, for pure information states, fidelity F does not depend on information state. This is also in contrast with the case of general pure entangled state resource. For $\gamma = 1$ (pure information states), the MASFI is $(1+\varepsilon)/2$. Variation of MASFI with parameters $\gamma$ and $\varepsilon$ is shown in fig.(1).

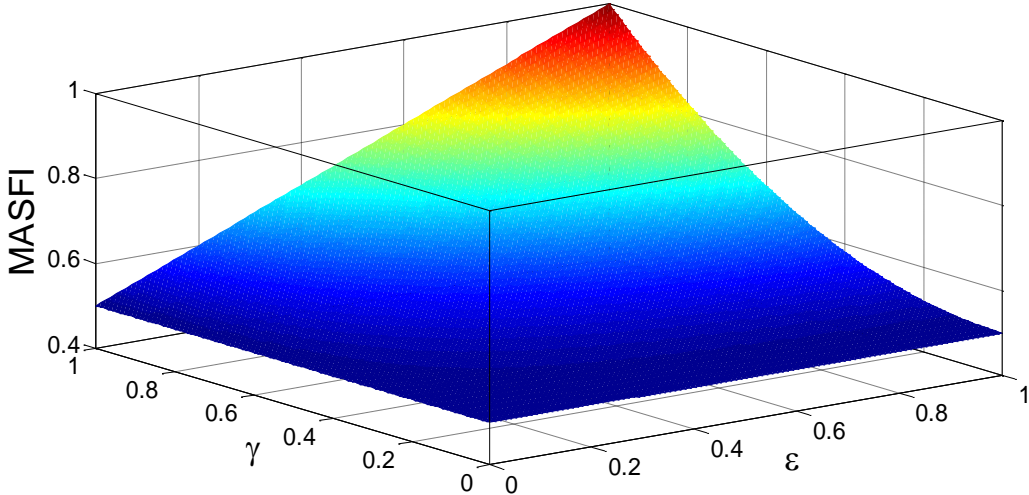

Fig. (1): variation of minimum assured fidelity (MASFI) with parameters $\gamma$ and $\varepsilon$.

For general mixed information state, the maximum possible value of fidelity is

$$F_{max} = \frac{1}{2}[1+\varepsilon], \qquad (15)$$

obtained at $(\theta,\phi,\alpha) = (0,0,0), (0,0,\pi), (0,\pi,0), (0,\pi,\pi)$. If one studies average of F over information state [5] and maximizes with respect to $\chi, \theta, \phi$ and $\psi$, the angles involved in unitary transformation, the result is

$$F_{av.max} = \frac{1}{2} + \frac{\varepsilon}{6}(1+2\gamma^2) \qquad (16)$$

at $(\theta,\phi) = (0,0)$. Clearly, MASFI $\leq F_{av.max} \leq F_{max}$. For pure information all three fidelities are identical and equal to $(1+\varepsilon)/2$, which is obvious from the fact that F ceases to depend on information when $\theta = \phi = 0$ is taken. It should be noted that if Alice does nothing and Bob chooses his unitary transformation randomly the average fidelity is 1/2 because for any randomly chosen state average probability of this state being in input state or in a state orthogonal to input state are equal.

MASFI is seen $\geq 1/2$ for all values of γ and ε. We note that $F_{av.max}$ is greater than classical fidelity (i.e., $F_{av.max} > \frac{2}{3}$) for $\varepsilon \geq (1+2\gamma^2)^{-1} \geq 1/3$, for which the Werner state is entangled with concurrence C = max{0, (3ε-1)/2}. However, this behavior is not shared by MASFI which is $\geq 2/3$ only if $\varepsilon \geq (3\gamma^2)^{-1}$ i.e., $C = (1-\gamma^{-2})/2$. MASFI is also $\geq 1/2$ for all values of ε and γ. We note that $F_{av.max}$ is $\geq 1/2$ for all cases of ε and γ. Behavior of $F_{av.max}$ is shown in Fig.(2).

We study the difference of MASFI and $F_{av.max}$ against ε and γ. From Eq.(15) and (17) we see that $F_{av.max} - MASFI = \frac{1}{6}(1-\gamma^2)\varepsilon > 0$, in general, the difference being zero at ε = 0 or at γ = 1. This difference increases with ε and decreases with γ. Thus, this variation is larger for lesser purity and for larger concurrence. The variation of $F_{av.max}$ - MASFI against $\gamma$ and $\varepsilon$ is shown in Fig. (3).

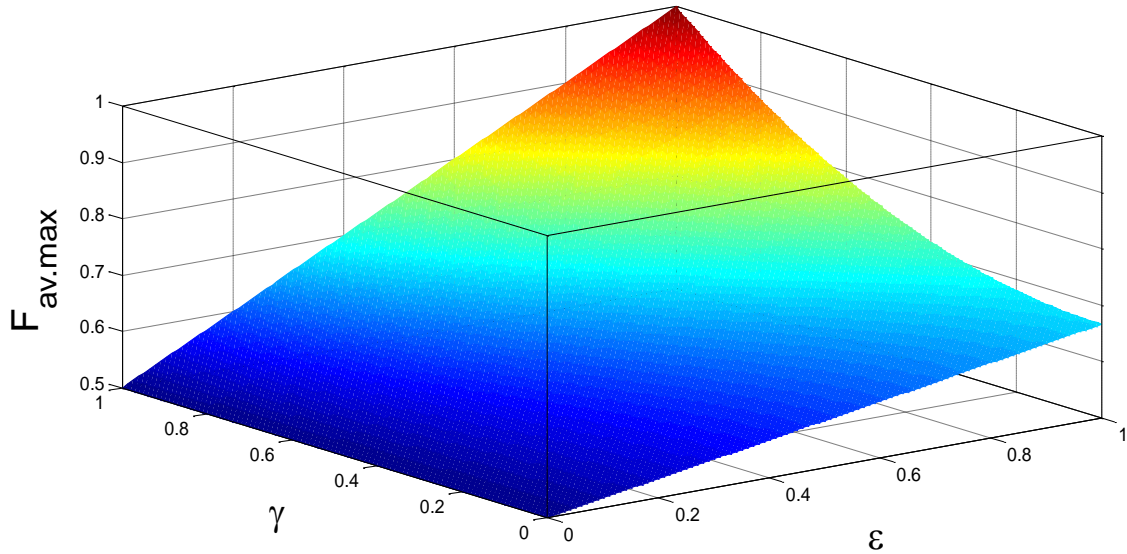

Fig. (2): variation of maximal average fidelity with parameters $\gamma$ and $\varepsilon$.

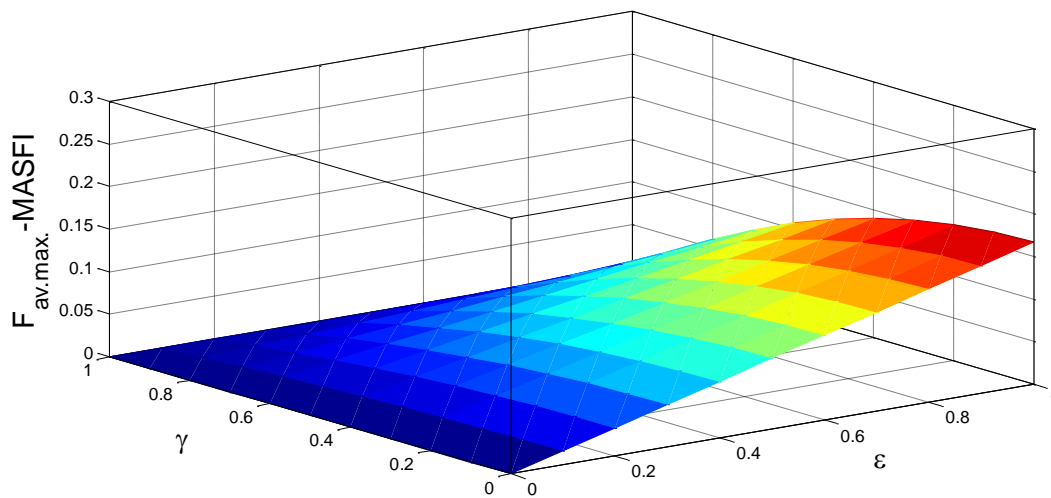

Fig. (3): The plot of difference between $F_{av.max}$ and MASFI against $\gamma$ and $\varepsilon$.

**Acknowledgements**

We are thankful to Prof. N.Chandra and Prof. R. Prakash for their interest and critical comments and to Dr.R.Kumar, Dr.P.Kumar, Dr.D.K. Mishra, A.K.Yadav, A.K. Maurya, and M. K. Mishra for helpful and stimulating discussions.

**References:**

[1] C.H.Bennett , G.Brassard , C Crepeau , R. Jozsa , A. Peres, and W.K. Wootters, Phy. Rev. Lett**. 70** (1993) 1895-1899.

[2] D.Bouwmeester. J.W.Pan, K.Mattle,M.Ebil,H.Weinfurter, and A Zeilinger, Nature (London) **390** (1997) 575.

[3] D.Boschi, S.Branca, F.De Martini, L. Hardy, and S. Popescu, Phy. Rev. Lett. **80** (1998) 1121.

[4] A. Furusawa, J.L. Sorensen, S.L. Braunstein, C.A Fuchs, H.J.Kimble, and E.S. Polzik, Science **282** (1998) 706.

[5 ] van Enk and Hirota, Phys. Rev. A 64 (2001) 022313.

[6] H. Prakash, N. Chandra, R. Prakash and Shivani, J. Phys. B: At. Mol. Opt. Phys. **40** (2007) 1613-1626 ; Phys. Rev. A **75** (2007) 044305-044308, also published in Vir. J. Quantum Inf., Vol.7, Issue 5, May (2007).

[7] A. Karlson, M. Bourennane; Phys. Rev. A 58 (1998) 4394.

[8] F.L. Yan, D. Wang; Phys. Lett. A 316 ( 2003) 297.

[9] Y. M. Li, K.S. Zhang and K.C.Peng; Phys. Lett A 324 (2004) 420.

[10] C.P. Yang, S-I. Chu and S. Han ; Phys. Rev A 70 (2004) 022329.

[11] F.G. Deng, C-Y. Li, Y-S. Li, H-Y. Zhou and Y. Wang; Phys. Rev. A 72 (2005) 022338.

[12] Y.Y. Nie, Z-H. Hong, Y-B. Huang, X-J. Yi and S-S Li; Int. J. Theor. Phys. 48 (2009) 1485.

[13] J. Dong and J.F. Teng; Eur.Phys. J. D 49 (2008) 129.

[14] C.P. Yang and G.C. Guo, Chin. Phys. Lett**. 17** (2000) 162.

[15] J. Lee, H. Min, and S.D. Oh, Phys. Rev. A **66** (2002) 052318.

[16] Gustavo Rigolin, Phy. Rev. **A 71** (2005) 032303.

[17] Fu-Guo Deng ,Phys. Rev. A **72** (2005) 036301.

[18] Chen P X, Zhu S Y and Guo G C; Phys. Rev. A **74** (2006) 032324.

[19] Man Z-X, Xia Y-J, An N B; Phys. Rev. A **75** (2007) 052306.

[20] Quan J N, Jian W Y; Chin. Phys. Lett. **27** (2010) 010302.

[21] Ming L D, Y-W. Wang, X-M. Jiang and Y-Z. Zheng; Chin. Phys. B **19** (2010) 020307.

[22] Qin Z X, Min L Y, Yun Z Z, Wen Z and Jun Z Z; Science China **53** (2010) 2069.

[23] Prakash H, Chandra N, Prakash R and Dixit A; MPLB **21** (2007) 2019.

[24] H.W. Lee, Phys. Rev. A **64** (2001) 014302.

[25] H. Prakash and V. Verma; arXiv preprint arXiv:1110.1220; H. Prakash et. al.; Mod. Phys. Lett. B **21** (2007) 2019.

[26] S. Hill and W.K. Wootters; Phys. Rev. Lett. **78** (1997) 5022.

[27] Valerie Coffman, Joydip Kundu, and William K. Wootters, Phy. Rev. A **61** (2000) 052306.

[28] H. Prakash, A.K. Maurya; Optics Communications 284 (2011) 5024.

[29] Hari Prakash and Vikram Verma; J. Phys. A: Math. Theor. 45 (2012) 395306.

[30] P. Agrawal and A.K. Pati; Phys. Lett. A 305 (2002) 12.

[31] A.K. Pati and P. Agrawal; J.Opt. B: quantum semi class Opt. 6 (2004) S844; Phys. Lett A 371 (2007) 185.

[32] H. Prakash, N. Chandra, R. Prakash and Shivani; Int. J. Mod. Phys. B 24 (2010) 3383; Shivani A. Kumar, H. Prakash, N. Chandra and R. Prakash; J. Quant. Inf. Science 2 (2012) 123.

[33] Hari Praksh and Vikram Verma *Quantum Inf Process*, **11** (2012) 1251.

[34] Hari Prakash and Manoj K. Mishra; J.Opt. Soc. Am. B 29(2012) 2915; arXiv:1107.2533 [quant-ph]; Manoj K Mishra and Hari Prakash;. J. Phys. B: At. Mol. Opt. Phys. 43 (2010) 185501.

[35] R.F. Werner, Phys. Rev. A **40** (1989) 4277.

[36] S. Ghosh, G. Kar, A. Sen. And U. Sen; Phys. Rev. A **64** (2001) 044301.

[37] T.C. Wei, K. Nemoto, P.M. Goldbart, P.G. Kwiat, W.J. Munro and F. Verstraete; Phys. Rev. A **67** (2003) 022110.